# CONSECUTIVE BRIGHT PULSES IN THE VELA PULSAR

Jim L Palfreyman[1], Aidan W Hotan[2], John M Dickey[1], Timothy G Young[3], and Claire E Hotan[1,4]

[1] School of Maths and Physics, University of Tasmania, Private Bag 37, Hobart, Tasmania 7001, Australia; jimp@postoffice.utas.edu.au
[2] CSIRO Astronomy and Space Science, PO Box 76, Epping NSW 1710, Australia
[3] Curtin Institute of Radio Astronomy, Curtin University of Technology, GPO Box U1987, Perth, Western Australia 6845, Australia
[4] CSIRO Marine and Atmospheric Research, GPO Box 1538, Hobart Tasmania 7001, Australia

## ABSTRACT

We report on the discovery of consecutive bright radio pulses from the Vela pulsar, a new phenomenon that may lead to a greater understanding of the pulsar emission mechanism. This results from a total of 345 hr worth of observations of the Vela pulsar using the University of Tasmania's 26 m radio telescope to study the frequency and statistics of abnormally bright pulses and sub-pulses. The bright pulses show a tendency to appear consecutively. The observations found two groups of six consecutive bright pulses and many groups of two to five bright pulses in a row. The strong radio emission process that produces the six bright pulses lasts between 0.4 and 0.6 s. The numbers of bright pulses in sequence far exceed what would be expected if individual bright pulses were independent random events. Consecutive bright pulses must be generated by an emission process that is long lived relative to the rotation period of the neutron star.

## 1. INTRODUCTION

The Vela pulsar (PSR J0835-4510) is the brightest pulsar in the southern sky and sits within the Vela supernova remnant at a distance of approximately 290 pc (Pellizzoni et al. 2010); it has a period of 89.3 ms, is highly linearly polarized and its rotation rate speeds up or "glitches" every few years (Lorimer & Kramer 2005, p. 49). Vela is not known to emit giant pulses of the kind typically associated with the Crab pulsar; however micro-giant pulses that are very narrow (~50 to ~400 μs) and have a peak flux density 40 times the integrated pulse have been observed. Vela's mean pulse envelope has a width of approximately 5 ms at 1413 MHz (Johnston et al. 2001). Vela exhibits a continuous distribution of pulse energies at radio wavelengths, although the shape of the energy spectrum has been observed to change as a function of phase angle or longitude (Cairns et al. 2003). Of particular interest is the leading edge of the main pulse window, where the energy spectrum has a power-law distribution with an extended tail of high-energy events. This study has the advantage of hundreds of hours of observations over a long time frame, providing a large sample of high-resolution single pulse flux measurements that expand our understanding of Vela's most energetic pulses.

A strict naming convention for different single pulse fluxes has not been formalized. Giant pulses are typically defined as being at least 10 times the flux of the average pulse integrated over the entire pulse window (Johnston et al. 2001). A "giant micro pulse" (Johnston et al. 2001) is a pulse that has a high peak flux density and typically has an FWHM of under 500 μs. A "sub-pulse" is a single pulse that is narrower than the mean pulse envelope. A "drifting sub-pulse" is a sequence of sub-pulses which drift monotonically in longitude with each successive pulse (Edwards and Stappers 2002). We define a "consecutive sub-pulse" to be a sub-pulse that occurs from one rotation to the next but does not appear as a permanent feature in the mean pulse profile. We also define a "bright" pulse as one that is at least five times the flux of the average pulse integrated over the entire pulse window, half the threshold for a true giant pulse but significantly greater than the average. Consequently, a "bright sub-pulse" is one that is both bright and a sub-pulse.

## 2. OBSERVATIONS AND DATA REDUCTION

Individual pulses from the Vela pulsar were observed from 2007 September through 2011 January with the University of Tasmania's 26 m radio telescope at Mt. Pleasant, near Hobart, Australia. The centre frequency of these observations is 1440 MHz with bandwidth 64 MHz. Typically a few hours of observations were made each week. The receiver consists of a cooled 20 cm prime-focus feedhorn with dual linear polarisation feeds that are sampled by an analog to digital converter using 2 bit precision at 128 million samples per second that were recorded directly onto a Redundant Array of Independent Disks. Processing was done using the software packages DSPSR and PSRCHIVE (Hotan et al. 2004) which perform phase-coherent de-dispersion using 16 frequency channels and 8192 pulse phase bins over the 89.3 ms pulse period. This gives a time resolution of 10.9 μs. We have collected over 300 hr of observations covering over $10^7$ pulses. A search of the data was conducted looking for bright pulses with a threshold of five times the average pulse strength. A second search was then conducted for consecutive bright pulses. Occasional bursts of interference (mainly due to overhead aircraft) were visually identified and all data taken during the interference episode were rejected. No sources of interference are seen with dispersion like that of the Vela pulsar, so we are confident that the bright pulses seen are not influenced by spurious signals.

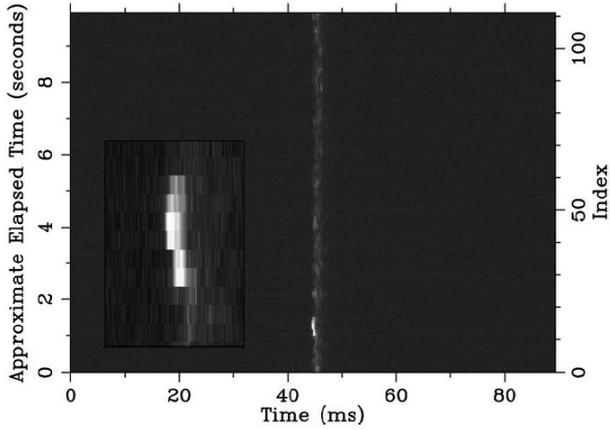

**Figure 1.** Plot of peak flux density vs. phase vs. time. The lighter the pixel the brighter the pulse. Including magnified view.

**Table 1**
Expected Consecutive Pulses vs. Observed

| Consecutive Pulses (n) | 2009 December Number Expected (q) | 2009 December Number Actually Observed | 2010 Number Expected (q) | 2010 Actually Observed |
|---|---|---|---|---|
| 2 | 133 | 191 | 30 | 578 |
| 3 | 0.74 | 70 | 0.17 | 27 |
| 4 | $4.1 \times 10^{-3}$ | 26 | $0.92 \times 10^{-3}$ | 5 |
| 5 | $23 \times 10^{-6}$ | 10 | $5.1 \times 10^{-6}$ | 1 |
| 6 | $127 \times 10^{-9}$ | 1 | $28 \times 10^{-9}$ | 0 |

This ongoing monitoring was followed up by an intense set of observations for 8 hr per day for 13 consecutive days for a total 108 hr of data starting 2009 day 350.

### 3. RESULTS

Figure 1 shows the first and one of the most interesting discoveries of this search so far: a series of six consecutive bright sub-pulses that started at 2008 June 29 07:32:20.9823 UTC. This figure is a time versus time graph folded at the period of the pulsar – the lighter the pixel the higher the flux density. The normal pulses can be seen as a faint vertical strip centred around the 45 ms mark. The bright sub-pulses are shown starting approximately one second from the bottom. A magnified view of these pulses is shown as an inset.

Figure 2 shows flux density vs time for each of these six bright sub-pulses with high resolution in pulse longitude.

Consecutive bright sub-pulses have not been discussed in the literature. They may give an indication of the nature of the pulsar emission process. The fact that they can last a number of pulsar rotations and in this case a total time of 0.5 s is significant. The six bright sub-pulses all occur at a similar longitude, in the leading edge of the pulse (as nearly all bright pulses in Vela do), and they show similar profiles, with a sharp leading edge and more gradual decrease after the peak.

Events with timescales around 0.5 s have not been previously noted in the Vela pulsar. Consecutive bright pulses of lesser length (i.e. 2, 3, 4 or 5 bright pulses in a row) were also noted in all data sets and were much more frequent (see Table 1).

Interstellar scintillation could cause amplification of a pulsar signal over short time periods. The NE2001 electron density model (Cordes & Lazio 2002) in the direction of the Vela pulsar predicts pulse broadening of 1.2 μs, a scintillation bandwidth of greater than 30 kHz and a scintillation timescale around 30 s. Since the expected time scale of amplification by scintillation is 60 times longer than the sequence of bright pulses, it is very unlikely that the bright pulses are an effect of propagation of the signal through the interstellar medium. The bright pulse sequences must be intrinsic to the pulsar emission mechanism or propagation through the pulsar's own magnetospheric plasma.

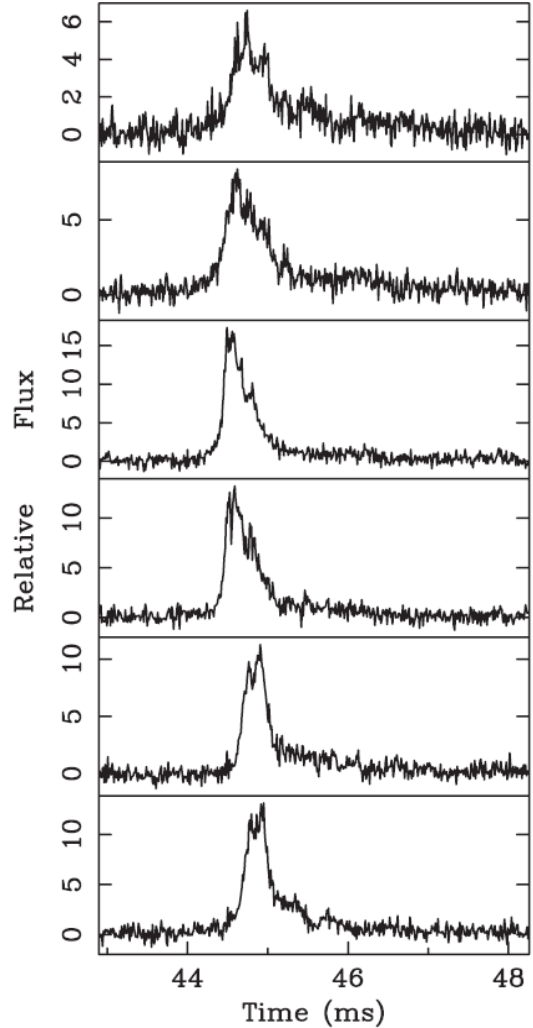

**Figure 2.** Plot of flux density vs. phase for each of the consecutive bright pulses. Note the y scale varies and the pulses are in the same order as Figure 1. The average pulse profile has a peak around 1 on this scale and a width of about 5 ms.

The random chance of consecutive bright sub-pulses occurring if there were no correlation in the emission process can be easily calculated. A single bright sub-pulse occurs in the 2009 December observations with a probability p=0.0053 (once every 190 pulses). Labelling n as the number of consecutive pulses and m the total number of pulses observed, the expected number q of n consecutive bright pulses, if they are uncorrelated, is approximated by

$$q \sim = mp^n \quad . \qquad (1)$$

Results for the number of consecutive bright pulses in the 108 hr of observations from the 13 day campaign in 2009 December and the total of 24.22 hr observed in 12 sessions in 2010 are given separately on Table 1.

In the 2009 December results, 191 sequences of two consecutive bright pulses are detected. This is notably higher than the prediction of Equation (1), which is 133. For n=3 we expect to see fewer than one sequence in 108 hr of observations, but we find 70. With n=4, 5, 6 there are several sequences detected, although the expected number based on Equation (1) is very small (Columns 2 versus 3 and 4 versus 5, Table 1).

Results of the 24.22 hr of observations in 2010 are shown in Columns 4 and 5 in the table. Another sequence of six consecutive bright pulses was not detected, but the shorter sequences of consecutive bright pulses were detected in numbers consistent with the results of the 2009 December campaign. The ratio of bright pulses to all pulses was 1/190 for the 2009 December observations and during the first half of 2010.

Vela "glitched" (sped up) on 2010 July 31 (Buchner, 2010). After the glitch the bright pulse rate increased to 1/111. This change in bright pulse statistics after the glitch, and the variation with time of the pulse height distribution will be discussed in a subsequent paper.

To see if the bright pulse statistics were sensitive to the arbitrary choice of threshold (five times the mean flux density), we re-calculated the statistics using a lower threshold. As expected, many more "bright" pulses and consecutive bright pulses were found with a lower threshold and the probabilities were still many orders of magnitude under what was observed.

Our conclusion is that the hypothesis that bright pulses are independent events is ruled out. The high incidence of bright pulse sequences shows that the physical process causing bright pulses is long-lived relative to the pulsar rotation rate.

## 4. DISCUSSION

It is generally accepted that a cone shaped beam centered on the magnetic axis of a pulsar accounts for many of the observations of a pulsar's pulse profile. The line of sight trajectory cuts the emission cone giving us a view of a curved slice of this beam (Lorimer & Kramer 2005, p67). There are a number of different hypotheses regarding the actual structure of this beam. The two main competing models are the nested cone structure (Rankin 1993) and the patchy beam structure (Lyne & Manchester 1988). These models are generally focused on the emission of the standard pulse and do not take into account the nature of bright or giant pulses.

Karastergiou & Johnston's (2007) empirical model for beams modifies the patchy beam structure by having different emission heights and different numbers of emitting patches per height. According to their model, older and slower pulsars tend to have many emission heights and a young pulsar like Vela would have a single emission height and about ten emission patches at that height. The consecutive bright pulses observed from Vela do not fit well with this model in its current form.

Kontorovich (2009) hypothesizes that electromagnetic energy is trapped in the cavity between the neutron star surface and the magnetosphere. The standard pulse is this energy: "radiating through a 'waveguide' near the magnetic axis or through a 'slot' on the border of the open field lines." Occasional breaks in the magnetosphere letting through this electromagnetic energy cause bright or giant pulses to be observed.

This inverts the traditional view in that the bright pulse reflects the normal state of the radiation field near the neutron star surface. The average pulse corresponds to strong attenuation through the magnetosphere. The observation that the consecutive bright pulses on Vela hold their phase position (before the main pulse) for several rotation periods indicates that these bright pulses are related to a specific part of the magnetosphere. Kontorovich's hypothesis is that these are gaps in the magnetosphere; in this context our observations would require that these gaps can remain open for a number of pulsar rotations.

Also of note is the fact that the last few pulses in the observation in Figure 2 are significantly lower in amplitude. This could be explained by the gap in the magnetosphere gradually closing.

Deshpande and Rankin (1999) study drifting sub-pulses in B0943+10 and produce a polar emission map showing, in a rotating frame, twenty emission centers around the magnetic axis. The pattern of emission centers takes 37 pulsar periods to complete one cycle around the magnetic axis. B0943+10 may provide a clue to the nature of the consecutive bright pulses observed in Vela, even though Vela does not have drifting sub-pulses. The period of B0943+10 is 1.098 s and the twenty emission centers are not tied to either the surface of the neutron star or to the magnetic field lines, and yet they remain stable over hundreds of seconds. If there is an unseen analog on Vela it would most likely rotate an order of magnitude faster than the one on B0943+10.

Combining the ideas proposed by Kontorovich and the stability and comparatively slow moving emission centers of Deshpande and Rankin, may lead to a refined emission model that explains the consecutive bright sub-pulse observations reported here. The distribution of lifetimes of magnetospheric windows, or of emission columns, may follow from these results. Searches of other young pulsars for similar sequences of bright sub-pulses would be very worthwhile, in order to determine whether Vela is peculiar or typical in showing this phenomenon. A change in the statistics of multiple bright sub-pulses after the glitch in late 2010 is the topic of continuing observations.

## 5. REFERENCES

Buchner, S. 2010, ATel, 2768
Cairns, I.H., Johnston, S., & Das, P. 2003, MNRAS 343, 512
Cordes, J. M., & Lazio, T. J. W. 2002, NE2001. I. A New Model for the Galactic Distribution of Free Electrons and its Fluctuations; arXiv:astro-ph/0207156
Deshpande, A. A., & Rankin, J. M. 1999, ApJ, 524:1008
Edwards, R. T., & Stappers, B. W. 2002, A&A 393, 733
Hotan, A. W., van Straten, W., & Manchester, R. N. 2004, PASA 21 (2004) 302
Johnston, S., van Straten, W., Kramer, M., & Bailes, M. 2001, ApJ, 549:L101
Karastergiou, A., & Johnston, S. 2007, MNRAS 380, 1678
Kontorovich, V. M. 2009, in Proc. The 8th International Conf. on Physics of Neutron Stars in Saint-Petersburg, 2008, ed. D.A. Varshalovich et al. (Saint Petersburg: Saint Petersburg State Polytech. Univ. Publishing), arXiv:0911.3272v2
Lorimer, D. R. & Kramer M. 2005, Handbook of Pulsar Astronomy (Cambridge Univ. Press)
Lyne A. G., & Manchester, R. N. 1988, MNRAS, 234, 477
Pellizzoni, A., et al. 2010, Science, 327, 663
Rankin, J. M., 1993, ApJ, 405, 285